\documentclass{iau-JDSS}
\usepackage{graphicx}

\title[Nuclear Star Clusters] 
{Nuclear Star Clusters\\ Structure and Stellar Populations}

\author[Nadine Neumayer]   
{Nadine Neumayer}

\affiliation{ESO, Karl-Schwarzschild Str. 2, 85748 Garching bei M\"unchen, Germany\\
email: {\tt nneumaye@eso.org}}

\pubyear{2012}
\volume{Volume 16}  
\pagerange{1--2}
\date{?? and in revised form ??}
\jname{Highlights of Astronomy, Volume 16}
\editors{Enrico Vesperini and Giampaolo Piotto, eds.}
\begin{document}

\maketitle

\begin{abstract}
This is an overview of nuclear star cluster observations, covering their structure, stellar populations, kinematics and possible connection to black holes at the centers of galaxies.
\end{abstract}

\firstsection 
\section{Nucleation Fraction, Stellar Populations, and Kinematics}
Nuclear Star Clusters (NCs) are a very common structural component at the centers of galaxies. They are found in $77\%$ of late type galaxies (\cite{boeker02}), 
$55\%$ of spirals (\cite{carollo98}), and at least $66\%$ of (dwarf) ellipticals and S0s (\cite{cote06}). These studies find that the half-light radii of NCs are typically $5pc$, and due to these small sizes, their detection requires very high spatial resolution observations, making the HST crucial for
systematic searches. NCs are on average 4mag brighter than GCs (\cite{boeker04}), i.e. they are more massive but have similar half-light radii. This makes NCs the densest stellar systems in the universe (\cite{walcher05,misgeld11}). They lie at the high-mass end of the star cluster mass function, and are structurally very different from bulges.

Nuclear star clusters truly occupy the centers of galaxies, both photometrically but also kinematically 
(\cite{boeker02, neumayer11}, respectively), and it may be this special location at the bottom of the potential well of the galaxies, that causes the star
formation history of NCs to be rather complex. Several studies have shown that NCs have multiple stellar populations both in late type (\cite{walcher06,seth06,rossa06}) and also early type galaxies 
(\cite{seth10}). The NC seems to be typically more metal-rich and younger than the surrounding galaxy (\cite{koleva11}), and the abundance ratios 
[$\alpha$/Fe] show that NCs are more metal enriched than globular clusters (GCs) (\cite{evstigneeva07}). This finding suggests that NCs cannot solely be the merger product of GCs, but need some gas for recurrent star formation. This finding is also supported by recent kinematical studies 
(\cite{hartmann11,delorenzi12}), where cluster infall alone cannot explain the dynamical state of the NC.

Recent studies of the kinematics of NCs with integral-field spectroscopy show that the cluster as a whole rotates (Seth et al. 2008, 2010). Combined with the superb spatial resolution of adaptive-optics, the 2D velocity maps resolve stellar and gas kinematics down to a few parsecs on physical scales. In addition, due to the extremely high central stellar density in NCs, it becomes possible to pick-up kinematic signatures for intermediate-mass black hole inside NCs (\cite{seth10}, Neumayer et al. in prep). 

\section{Connection to Black Holes}
\begin{figure}[]
  \begin{minipage}[c]{0.52\textwidth}
    \includegraphics[width=\textwidth]{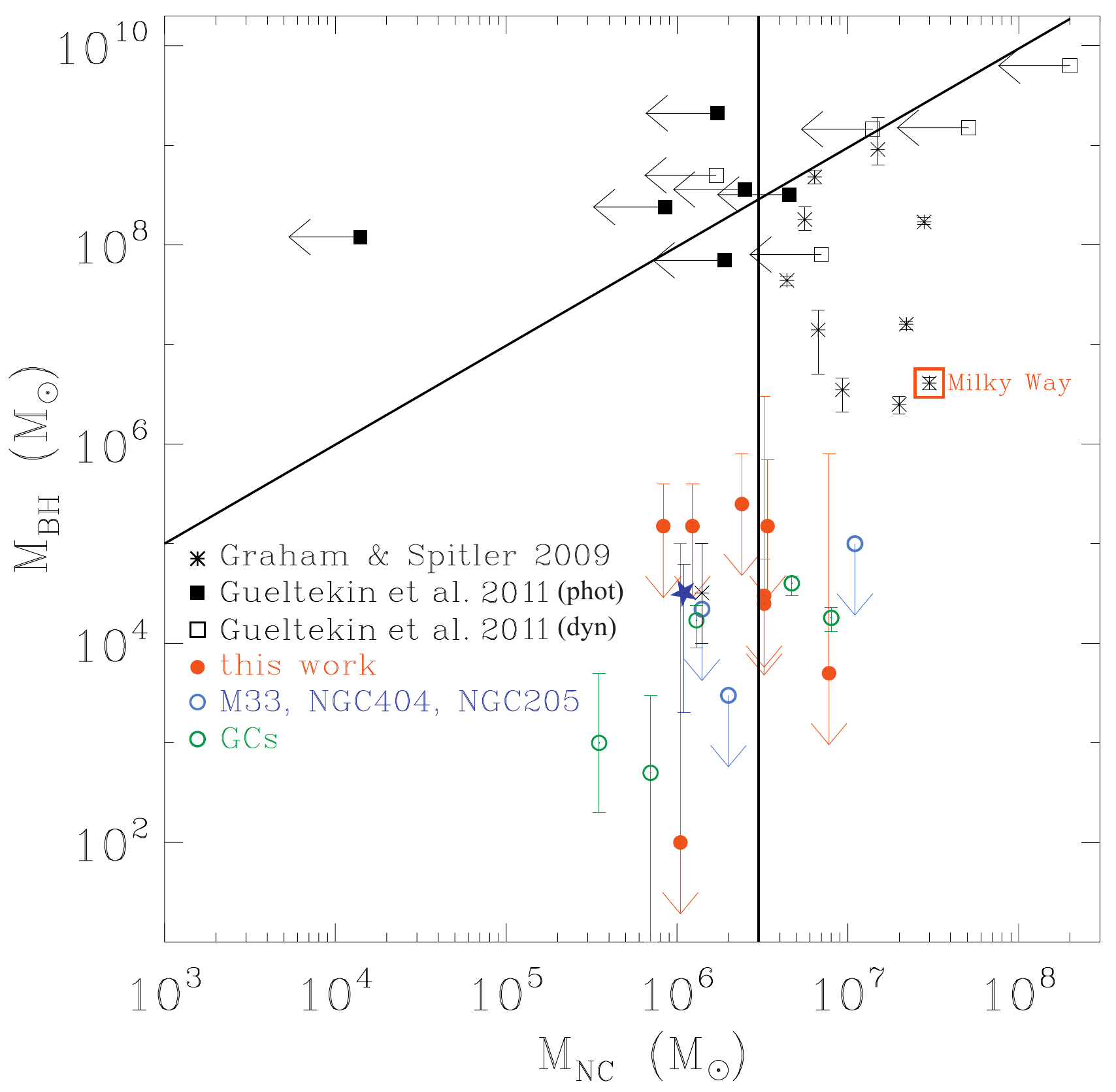}
  \end{minipage}\hfill
  \begin{minipage}[c]{0.45\textwidth}
    \vspace{1.8cm}
    \caption{
       {The mass of the BH mass vs. the NC mass. The two full lines indicate a NC
	mass of $3 \times 10^6 M_{\odot}$ and a MBH / MNC mass ratio of 100. These
	lines separate NC dominated galaxy nuclei (lower left of both lines)
	from BH dominated galaxy nuclei (upper left of both lines) and a
	transition region (to the right of both lines) (Figure~5 from \cite{neumayer12}).
    } \label{fig:MBH_MNC}}
  \end{minipage}
\end{figure}

Unlike black holes, NCs provide a visible record of the accretion of stars and gas into the center of a galaxy, and studying their stellar populations, structure and kinematics allow us to disentangle their formation history. 
NCs do co-exist with black holes. The best studied example is the NC in our own Galaxy (see R. Sch\"odel this edition). But there are also NCs with very tight upper limits on the mass of a central black hole (see \cite{neumayer12} for an overview), and it is not yet clear under what conditions galaxies make NCs and/or BHs. Figure~\ref{fig:MBH_MNC} shows a compilation of mass measurements of BHs and NCs. For the lowest mass NCs BHs are very hard to detect (if present), while for the highest mass BHs, the surrounding NCs seem to have been destroyed already. The underlying physical connection remains unclear for now, but it may well be that BHs grow inside NCs, that may thus be the precursors of massive BHs at the nuclei of galaxies.

\begin{acknowledgments}
I acknowledge the support by the DFG cluster of excellence `Origin and Structure of the Universe', and thank the IAU for financial support.
\end{acknowledgments}

\end{document}